\documentclass[aps,prl,twocolumn,showpacs,groupedaddress]{revtex4}
\usepackage{epsfig}
\usepackage{dcolumn}   
\usepackage{bm}        
\usepackage{amssymb}

\begin{document}
\title{Fast Equilibration of Hadrons in an Expanding Fireball}

\author{J. Noronha-Hostler$^{1}$}
\author{C. Greiner$^{2}$}
\author{I. A. Shovkovy$^{3}$}
\affiliation{$^{1}$Frankfurt Institute for Advanced Studies, J. W.
Goethe Universit\"at, D-60438 Frankfurt am Main, Germany}
\affiliation{$^{2}$Institut f\"ur Theoretische Physik, J. W.
Goethe Universit\"at, D-60438 Frankfurt am Main, Germany}
\affiliation{$^{3}$Western Illinois University, Macomb, IL, 61455 USA}
\begin{abstract}

Due to long chemical equilibration times within standard hadronic reactions during the hadron gas phase in relativistic heavy ion collisions it has been suggested that the hadrons are ``born" into equilibrium after the quark gluon plasma phase.   Here we develop a dynamical scheme in which possible Hagedorn states 
contribute to fast chemical equilibration times of baryon 
anti-baryon pairs (as well as kaon anti-kaon pairs) inside a hadron gas and just below the critical temperature. Within this scheme, we use master equations and 
derive various analytical estimates for the chemical equilibration times. Applying a Bjorken picture to the expanding fireball, the kaons and baryons as well as the bath of pions and Hagedorn resonances can indeed quickly chemically equilibrate for both an initial overpopulation or underpopulation of Hagedorn resonances. Moreover, a comparison of our results to $(B+\bar{B})/\pi^{+}$ and $K/\pi^{+}$ ratios at RHIC, indeed, shows a close match.
\end{abstract}
\maketitle

(Anti-)strangeness enhancement was first observed at CERN-SPS energies by comparing anti-hyperons, multi-strange baryons, and kaons to $pp$-data.  It was considered a signature for quark gluon plasma (QGP) because, using binary strangeness production and exchange reactions, chemical equilibrium could not be reached  within the hadron gas phase \cite{Koch:1986ud}.  It was then proposed that there exists a strong hint for QGP at SPS because strange quarks can be produced more abundantly by gluon fusion, which would account for strangeness enhancement following hadronization and rescattering of strange quarks. Later, multi-mesonic reactions were used to explain secondary production of anti-protons and anti-hyperons \cite{Rapp:2000gy,Greiner}.
At SPS they give a typical chemical equilibration time $\tau_{\bar{Y}}\approx 1-3\frac{\mathrm{fm}}{c}$ using an annihilation cross section of $\sigma_{\rho\bar{Y}}\approx\sigma_{\rho\bar{p}}\approx 50\mathrm{mb}$ and  a baryon density of $\rho_{B}\approx \rho_{0}\;\mathrm{to}\;2\rho_{0}$, which is typical for SPS.
Therefore, the time scale is
short enough to account for chemical equilibration within a cooling hadronic
fireball at SPS.

A problem arises when the same multi-mesonic reactions were employed in the hadron gas phase at RHIC temperatures  where experiments show that the particle abundances reach chemical equilibration close to the phase transition \cite{Braun-Munzinger}.  At RHIC at $T=170$ MeV, where $\sigma\approx
30\mathrm{mb}$ and 
$\rho_{B}^{eq}\approx\rho_{\bar{B}}^{eq}\approx0.04\mathrm{fm}^{-3}$, the equilibrium rate for (anti-)baryon production is
$\tau\approx
10\frac{\mathrm{fm}}{\mathrm{c}}$, which is considerably longer than the fireball's lifetime in the hadronic stage of $\tau<5\frac{\mathrm{fm}}{\mathrm{c}}$.  Moreover, $\tau\approx
10\frac{\mathrm{fm}}{\mathrm{c}}$ was also obtained in Ref.\ \cite{Kapusta} using a fluctuation-dissipation theorem
and  a significant deviation  was found in the population number of various (anti-)baryons  from experimental data in the $5\%$ most central Au-Au collisions \cite{Huovinen:2003sa}. 
These discrepancies suggest that hadrons are ``born" into equilibrium, i.e.,
the system is already in a chemically frozen out state at the end of
the phase transition \cite{Stock:1999hm,Heinz:2006ur}.  

In order to circumvent such long time scales it was suggested that near $T_{c}$ there exists an extra large particle density overpopulated with pions and kaons, which drive the baryons/anti-baryons into equilibrium \cite{BSW}.  But it is not clear how this overpopulation should appear, and how the subsequent population of (anti-)baryons would follow.   Moreover, the overpopulated (anti-)baryons do not later disappear \cite{Greiner:2004vm}. Therefore, it was conjectured that Hagedorn resonances (heavy resonances near $T_{c} $ with an exponential mass spectrum) could account for the extra (anti-)baryons \cite{Greiner:2004vm}.  
Baryon anti-baryon \cite{Greiner:2004vm,Noronha-Hostler:2007fg} and kaon anti-kaon production develop according to
\begin{eqnarray}\label{eqn:decay}
n\pi&\leftrightarrow &HS\leftrightarrow n_{i,b}\pi+B\bar{B},\nonumber\\
n\pi&\leftrightarrow &HS\leftrightarrow n_{i,k}\pi+K\bar{K},
\end{eqnarray} 
which provide an efficient method for producing baryons and kaons because of the large decay widths of the Hagedorn states. In Eq. (\ref{eqn:decay}), $n$ is the number of pions for the decay $n\pi\leftrightarrow HS$, which can vary, and $n_{i,b}$ ($n_{i,k}$) is the number of pions that a Hagedorn state will decay into when a baryon anti-baryon (kaon anti-kaon) pair is present.
Since Hagedorn resonances are highly unstable, the phase space for multi-particle decays drastically increases when the mass is increased.  Therefore, the resonances catalyze rapid equilibration of baryons and kaons near $T_{c} $ where the Hagedorn states show up. Here we use a Bjorken expansion within a cooling fireball in order to see at which temperature the chemical equilibrium values are reached or maintained. In this letter we also briefly discuss an analytical solution of the chemical equilibration time, which is valid at a constant temperature near $T_{c}$. Moreover, our numerical results for the baryon anti-baryon pairs and kaon anti-kaon pairs suggest that the hadrons can, indeed, be born \textit{out} of  equilibrium.

We use a truncated Hagedorn mass spectrum \cite{Hagedorn:1968jf} 
\begin{equation}
g(m)=\int_{M_{0}}^{M}\frac{A}{\left[m^2 +(m_{0})^2\right]^{\frac{5}{4}}}e^{\frac{m}{T_{H}}}dm
\end{equation}
where the Hagedorn temperature is set to $T_{H}=180$MeV, which lies within the present range of Lattice QCD predictions \cite{Karsch}, the normalization factor is $A=0.5\;\textrm{MeV}^{\frac{3}{2}}$, and $m_{0}=0.5$ GeV.  We consider only mesonic, non-strange resonances and discretize the spectrum into mass bins of 100 MeV that range from the mass $M_{0}=2$ GeV to $M=7$ GeV.  The effects of the truncation and Hagedorn temperature are further discussed in \cite{NHCGISbig}.  However, the values we have chosen are acceptable.  

The abundances' evolution of the Hagedorn states,  pions, and baryon anti-baryon pairs due to the reactions in Eq.\ (\ref{eqn:decay}) are described by the following rate equations
\begin{eqnarray}\label{eqn:rate}
\dot{\lambda}_{i}&=&\Gamma_{i,\pi}\left(\sum_{n=2}^{\infty} B_{i, n}
\lambda_{\pi}^{n}-\lambda_{i}\right)+\Gamma_{i,B\bar{B}}\left( 
\lambda_{\pi}^{\langle n_{i,b}\rangle} \lambda_{B\bar{B}}^2 -\lambda_{i}\right),\nonumber\\
\dot{\lambda}_{\pi }&=&\sum_{i} \Gamma_{i,\pi}\frac{N_{i}^{eq}}{N_{\pi}^{eq}}  \left(\lambda_{i}\langle n_{i}\rangle-\sum_{n=2}^{\infty}
B_{i, n}n\lambda_{\pi}^{n} \right)\nonumber\\
&+&\sum_{i} \Gamma_{i,B\bar{B}} \langle n_{i,b}\rangle\frac{N_{i}^{eq}}{N_{\pi}^{eq}}\left(\lambda_{i}-
\lambda_{\pi}^{\langle n_{i,b}\rangle} \lambda_{B\bar{B}}^2\right),  \nonumber\\
\dot{\lambda}_{B\bar{B}}&=&\sum_{i}\Gamma_{i,B\bar{B}}\frac{N_{i}^{eq}}{N_{B\bar{B}}^{eq}}\left( \lambda_{i}- \lambda_{\pi}^{\langle n_{i,b}\rangle} \lambda_{B\bar{B}}^2\right),
\end{eqnarray}
where the fugacity is $\lambda=\frac{N}{N^{eq}}$, $N$ is the total number of each particle, and its respective equilibrium value is $N^{eq}$. The summation over $i$ represents the $i^{th}$ Hagedorn resonance bin. The structure of the rate equations for the kaon anti-kaon pairs is the same as in Eq. (\ref{eqn:rate}), however, $K\bar{K}$ is substituted in for $B\bar{B}$.  

The branching ratios for $HS\leftrightarrow n\pi$ are described by a Gaussian distribution
$B_{i, n}\approx
\frac{1}{\sigma_{i}\sqrt{2\pi}}e^{-\frac{(n-\langle n_{i}\rangle)^{2}}{2\sigma_{i} ^{2}}}$ 
where $\langle n_{i}\rangle=0.9+1.2\frac{m_{i}}{m_{p}}$ is the average pion number each
Hagedorn state decays into, found in a microcanonical model\cite{Liu}, $\sigma_{i}^{2}=(0.5\frac{m_{i}}{m_{p}})^{2}$ is the chosen width of the
distribution, and $n\geq 2$ is the cutoff for the pion number. Moreover, the branching ratios are normalized such that $\sum_{n=2}^{\infty} B_{i, n}=1$, which gives $\langle n_{i}\rangle\approx 2$ to $9$ and  $\sigma_{i}^{2}\approx 0.8$ to $11$.
The total decay width, $\Gamma_{i}\approx 0.17m_{i}-88$ MeV, which ranges from $\Gamma_{i}=250\;\mathrm{to}\;1100$ MeV, was found using the mass and decay widths in  \cite{Eidelman:2004wy} and fitting them linearly similarly to what was done in Ref.\ \cite{LizziSenda}.  The decay widths for the baryon anti-baryon
decay are $\Gamma_{i,B\bar{B}}=\langle
B\rangle\Gamma_{i}$ and $\Gamma_{i,\pi}=\Gamma_{i}-\Gamma_{i,B\bar{B}}$.  The average baryon number $\langle B\rangle $ per unit decay of Hagedorn resonances within a microcanonical model ranges from $0.06\mathrm{\;to}\;0.4$, so
$\Gamma_{i,B\bar{B}}=15\;\mathrm{\;to}\;400$ MeV \cite{Greiner:2004vm}. We use only the average values in Eq.\ (\ref{eqn:decay}) so that $\langle n_{i,b}\rangle=\langle n_{i,k}\rangle=3$ to 6 \cite{Liu,Greiner:2004vm} is used for both the baryons and kaons.  For the kaons $\Gamma_{i,K\bar{K}}=\langle K\rangle \Gamma_{i}$ where $\langle K\rangle=0.4$ to $0.5$ \cite{Liu,Greiner:2004vm}.  Thus, heavier resonances equilibrate more quickly because of large decay widths.  

Using a Bjorken expansion, we find a relationship between the temperature and the time, $T(t)$, for which the total entropy is held constant
\begin{equation}\label{eqn:constrain}
\mathrm{const.}=s(T)V(t)=\frac{S_{\pi}}{N_{\pi}}\int \frac{dN_{\pi}}{dy} dy
\end{equation}
where $\int \frac{dN_{\pi}}{dy} dy=874$ from Ref.\ \cite{Bearden:2004yx} for the $5\%$ most central collisions within one unit of rapidity and  the entropy per pion $S_{\pi}/N_{\pi}=5.5$ is larger than that for a gas of massless pions according to the analysis in Ref.\ \cite{Greiner:1993jn}.  The volume \cite{Greiner} is
\begin{equation}\label{eqn:bjorken}
V_{eff}(t\geq t_{0})=\pi\;ct\left[r_{0}(t_0)+v_{0}(t-t_{0})+\frac{a_{0}}{2}(t-t_{0})^2 \right]^2
\end{equation}
where the initial radius is $r_{0}(t_0)=7.1$ fm,  the average transversal velocity varies $v_{0}=0.3c,0.5c,$ and $0.7c$, and  the corresponding accelerations are taken as
$a_{0}=0.035,0.025,$ and $0.015$, respectively.  

\begin{figure}[h]
\centering
\begin{tabular}{c}
\epsfig{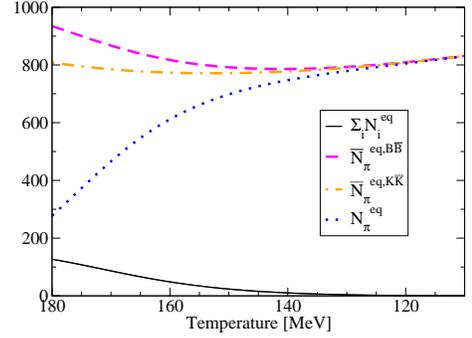}
\end{tabular}
\caption{Comparison of the total equilibrium number of pions $N_{pi}^{eq}$, Hagedorn states $\sum_{i}N_{i}^{eq}$, and effective pions $\tilde{N}_{\pi}^{eq}$ as defined in Eq.\ (\ref{eqn:effpi}).} \label{fig:density}
\end{figure}

The equilibrium values of pions, $N_{\pi}^{eq}$, shown in Fig.\ \ref{fig:density} are found using a statistical model \cite{StatModel}. 
Here we consider both the direct pions and the indirect pions, which come from resonances such as $\rho$, $\omega$ etc, and both the direct and indirect kaons. 
In Fig.\ \ref{fig:density} we see that $N_{\pi}^{eq}$ increases with decreasing temperature. This occurs because the Hagedorn states dominate the entropy at high temperatures, which affects $N_{\pi}^{eq}$ due to the entropy constraint in Eq.\ (\ref{eqn:constrain}).  Therefore, we must consider the number of ``effective pions" in the system, i.e., the total number of pions plus the potential number of pions from the Hagedorn resonances, defined as
\begin{eqnarray}\label{eqn:effpi}
\tilde{N}_{\pi,K\bar{K}}&=&N_{\pi}+\sum_{i}N_{i}\left(\frac{\Gamma_{i,\pi}}{\Gamma_{i}}\langle n_{i}\rangle +\frac{\Gamma_{i,K\bar{K}}}{\Gamma_{i}}\langle n_{i,k}\rangle\right)\nonumber\\
\tilde{N}_{\pi,B\bar{B}}&=&N_{\pi}+\sum_{i}N_{i}\left(\frac{\Gamma_{i,\pi}}{\Gamma_{i}}\langle n_{i}\rangle +\frac{\Gamma_{i,B\bar{B}}}{\Gamma_{i}}\langle n_{i,b}\rangle\right)
\end{eqnarray}
for the kaons and baryons, respectively.  In both cases $\tilde{N}_{\pi}^{eq}$ remain roughly constant throughout the Bjorken expansion. Additionally, throughout this paper our initial conditions are the various fugacities
$\alpha\equiv\lambda_{\pi}(t_0)$, $\beta_{i}\equiv\lambda_{i}(t_0)$, and $\phi\equiv\lambda_{B\bar{B}}(t_0)$ or $\phi\equiv\lambda_{K\bar{K}}(t_0)$, which are chosen by holding the contribution to the total entropy from the Hagedorn states and pions constant i.e. $s_{Had}(T_{0},\alpha)V(t_{0})+s_{HS}(T_{0},\beta_{i})V(t_{0})=s_{Had+HS}(T_{0})V(t_{0})=const$. 

The initial estimate for the Hagedorn state equilibration time is $\tau_{i}\equiv1/\Gamma_{i}$.  In order to estimate the chemical equilibration time, we use Eq.\ (\ref{eqn:rate}) in a static environment to find the equilibration time to be in the general ballpark \cite{Greiner,Greiner:2004vm} of
\begin{eqnarray}\label{eqn:taubbkk}
\tau_{B\bar{B}}&\equiv&\frac{N_{B\bar{B}}^{eq}}{\sum_{i}\Gamma_{i,B\bar{B}}N_{i}^{eq}}=0.2-0.7\;\frac{\mathrm{fm}}{c}\nonumber\\
\tau_{K\bar{K}}&\equiv&\frac{N_{K\bar{K}}^{eq}}{\sum_{i}\Gamma_{i,K\bar{K}}N_{i}^{eq}}=0.1-0.3\;\frac{\mathrm{fm}}{c}
\end{eqnarray}
between $T=180$ to $170$ MeV. 
As will be proven in detail in Ref.\ \cite{NHCGISbig}, these time scales are only precise when the pions and Hagedorn states are held in equilibrium.  In reality the chemical equilibration times are more complicated due to non-linear effects and the evolution of the equilibration must be divided into separate stages for a sufficient analysis.  

\begin{figure}
\centering
\epsfig{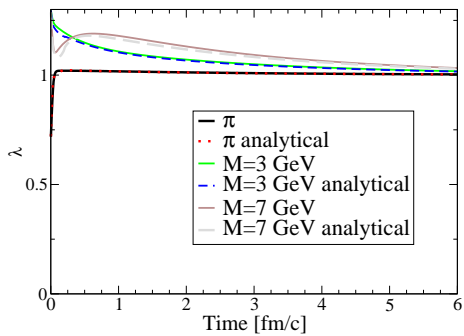}
\caption{Numerical and analytical results in a static environment at $T=180$ MeV when  $\beta_{i}=1.3$ and $\alpha=0.7$.} \label{fig:pifree}
\end{figure} 
\begin{figure}
\centering
\epsfig{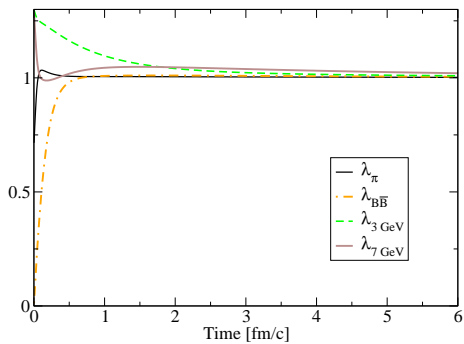}
\caption{Same as Fig.\ \ref{fig:pifree} with no initial baryons ($\phi=0$).} \label{fig:BBalso}
\end{figure} 
\begin{table}
\begin{center}
 \begin{tabular}{|c|c|c|}
 \hline
   & Time Scale & $T=180 - 170 $MeV \\
 \hline
 $\lambda_{\pi}\approx 0$ & $\tau_{\pi}^{0}\equiv\frac{N_{\pi}^{eq}}{\sum_{i} \Gamma_{i} N^{eq}_{i} \beta_{i}}$ & $0.1-0.4\;\frac{\mathrm{fm}}{c}$\\
 $\lambda_{\pi}\approx 1$ & $\tau_{\pi}\equiv\frac{N_{\pi}^{eq}}{\sum_{i} \Gamma_{i} N^{eq}_{i} \langle n_{i}^2\rangle}$ & $0.02-0.06\;\frac{\mathrm{fm}}{c}$\\
 QE & $\tau^{QE}_{\pi}\equiv\frac{N_{\pi}^{eq}}{\sum_{i} \Gamma_{i} N^{eq}_{i} \sigma_{i}^2}+\frac{\sum_{QE}N_{i}^{eq}\langle n_{i}^2\rangle}{\sum_{i}\Gamma_{i}N_{i}^{eq}\sigma_{i}^2}$ & $2.7-3.7\;\frac{\mathrm{fm}}{c}$\\
 \hline
 TOT & $\tau^{tot}\equiv\tau_{2GeV}+\tau^{QE}_{\pi}$ & $3.5-4.5\;\frac{\mathrm{fm}}{c}$\\
 \hline 
 \end{tabular}
 \end{center}
 \caption{Equilibration times from analytical estimates where QE is quasi-equilibrium and TOT is total equilibrium}\label{tab:tau}
 \end{table}
To find time scale estimates, we consider the more simplified case near $T_{c}$ excluding the baryons and kaons, i.e., Eq.\ (\ref{eqn:rate}) without the baryonic terms. The evolution of the rate equations can be divided into three stages as shown in Tab.\ \ref{tab:tau} and derived in \cite{NHCGISbig}. Initially, when the pions are far from equilibrium ($\lambda_{\pi}\approx 0$) the Hagedorn states can be held constant at a constant fugacity $\beta_{i}$.  Substituting $\lambda_{\pi}\approx 0$ and $\lambda_{i}\approx\beta_{i}$ into Eq.\ (\ref{eqn:rate}), we obtain $\tau_{\pi}^{0}$.  As the pions near equilibrium, we can then use $\lambda_{\pi}\rightarrow 1$ to obtain $\tau_{\pi}$.  Eventually, the right-hand sides of Eq.\ (\ref{eqn:rate}) become roughly zero before full equilibration (known as quasi-equilibrium), which occurs once the lightest resonance reaches quasi-equilibrium $\tau_{2GeV}=0.8\frac{fm}{c}$.  To obtain $\tau^{QE}_{\pi}$ we solved Eq.\ (\ref{eqn:effpi}) without the baryonic term, assuming $\lambda_{\pi}\rightarrow 1$ and that the right-hand side of the pion rate equation equals zero.  Then the total equilibration time $\tau^{tot}$ is just the addition of $\tau_{2GeV}$ and $\tau^{QE}_{\pi}$.  Since $\tau^{tot}$ includes all the non-linear effects, which occur even after equilibrium is neared, the more appropriate time scale is on the order of $\tau_{\pi}^{0}$. The equilibration times increase directly with $N_{\pi}^{eq}$, $\langle n_{i}^2\rangle$ and are shortened by large $\Gamma_{i}$'s and wide branching ratio distributions  $\sigma_{i}$'s. Because $N_{i}^{eq}$ decreases quickly as the system is cooled, the equilibration time is significantly longer at lower temperatures.  In Fig.\ \ref{fig:pifree} our analytical fit, which are exponential fits \cite{NHCGISbig} based on Tab.\ \ref{tab:tau},  match our numerical results well and nicely concur with the numerical results in Fig.\ \ref{fig:BBalso}.  Additionally, the baryons take slightly longer than predicted in Eq.\ (\ref{eqn:taubbkk}), but they still equilibrate quickly (Fig.\ \ref{fig:BBalso}).
\begin{figure}[t!]
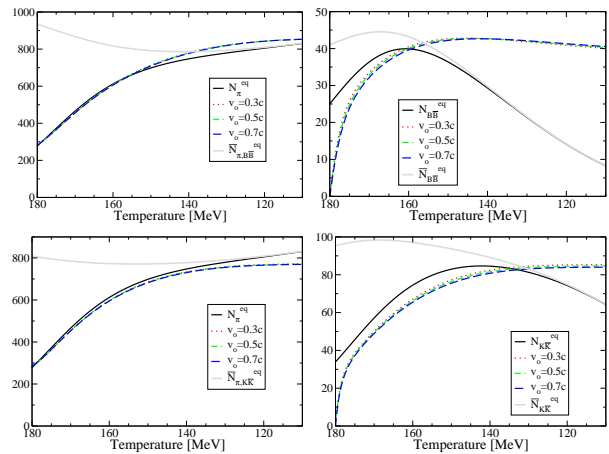

\centering
\begin{tabular}{cc}
\epsfig{file=EX_HS1.0_PI1.0_BB0.0_PI.eps,width=0.45\linewidth,clip=} &
\epsfig{file=EX_HS1.0_PI1.0_BB0.0_BB.eps,width=0.45\linewidth,clip=} \\
\epsfig{file=EX_HS1.0_PI1.0_KK0.0_PI.eps,width=0.45\linewidth,clip=} &
\epsfig{file=EX_HS1.0_PI1.0_KK0.0_KK.eps,width=0.45\linewidth,clip=} \\
\end{tabular}
\caption{Results for an expanding fireball when $\alpha=1$, $\beta=1$, and $\phi=0$. The effective number of baryons $\tilde{N}_{B\bar{B}}^{eq}$, kaons $\tilde{N}_{K\bar{K}}^{eq}$, and pions $\tilde{N}_{\pi,B\bar{B}}^{eq}$ and $\tilde{N}_{\pi,K\bar{K}}^{eq}$ are given.} \label{fig:expanshown}
\end{figure} 
\begin{figure}[t!]
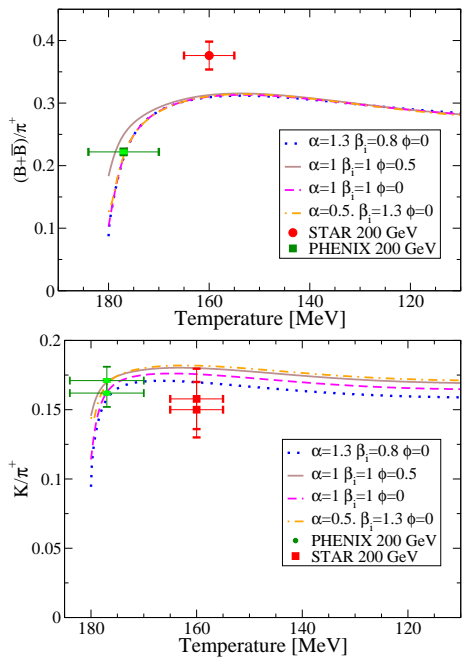

\centering
\epsfig{file=b2pi.eps,width=0.7\linewidth,clip=} \\
\epsfig{file=k2pi.eps,width=0.7\linewidth,clip=}
\caption{Comparison of the ratios obtained employing the expanding fireball picture for various initial conditions to data at $\sqrt{s}=200$ GeV from PHENIX \cite{PHENIX} and STAR \cite{STAR}.} \label{fig:expan}
\end{figure} 

In Fig.\ \ref{fig:expanshown} the baryons and kaons are shown for an expanding system where we see that the baryons reach chemical equilibrium by $T=165$ MeV ($t-t_{0}\approx2-3\;\frac{fm}{c}$) and the kaons at $T=160-140$ MeV. As with the pions, we consider the effective number of baryons and kaons because of the effects of Hagedorn resonance on the entropy at high temperatures, so
\begin{eqnarray}\label{eqn:effbbkk}
\tilde{N}_{B\bar{B}}&=&N_{B\bar{B}}+\sum_{i}N_{i}\frac{\Gamma_{i,B\bar{B}}}{\Gamma_{i}}\nonumber\\
\tilde{N}_{K\bar{K}}&=&N_{K\bar{K}}+\sum_{i}N_{i}\frac{\Gamma_{i,K\bar{K}}}{\Gamma_{i}},
\end{eqnarray}
which are shown in Fig.\ \ref{fig:expanshown}.  Not surprisingly, $\tilde{N}_{\pi,B\bar{B}}^{eq}$ and $\tilde{N}_{\pi,K\bar{K}}^{eq}$ remain almost constant due to the constraint set in Eq.\ (\ref{eqn:constrain}). Moreover, our expansion is not strongly affected by $v_{0}$ and, therefore, in the following graphs it is set to $v_{0}=0.5c$.

In Fig.\ \ref{fig:expan} we compare our total baryon to pion ratio $(B+\bar{B})/\pi^{+}$ to experimental data from PHENIX \cite{PHENIX} and STAR \cite{STAR}. $(B+\bar{B})/\pi^{+}$ is calculated by $B+\bar{B}=p+\bar{p}+n+\bar{n}\approx 2(p+\bar{p})$.  It should be noted here that in our calculations we use both the effective number of baryons, in Eq.\ (\ref{eqn:effbbkk}), and pions, in Eq.\ (\ref{eqn:effpi}). We obtain  $(B+\bar{B})/\pi^{+}\approx 0.3$, which matches the experimental data well.  Moreover, our results are independent of the chosen initial conditions. Also, in Fig.\ \ref{fig:expan} we compared the kaon to pion ratio  to the data at PHENIX \cite{PHENIX} and STAR \cite{STAR} (both $K/\pi^{+}$ and $\bar{K}/\pi^{+}$ are shown).   Again, we use the effective number of kaons (\ref{eqn:effbbkk}) and pions (\ref{eqn:effpi}). Our $K/\pi^{+}$ ratios compare to the experimental data very well and they level off between 0.16 to 0.17. As with the baryon anti-baryon pairs we do not see a very strong dependence on our initial conditions. In Fig.\ \ref{fig:expan} both figures agree well with experimental data.  Moreover, they remain roughly constant after $T=170-160$ MeV.  This demonstrates that the potential Hagedorn states can be used to explain dynamically the build up of the known particle yields.  

In future work we will consider strange baryonic degrees of freedom  and thoroughly study the effects of our initial conditions and parameters. 
To conclude, we used Hagedorn resonances as a dynamical mechanism to quickly drive baryons and kaons into equilibrium between temperatures of $T=165-140$ MeV.  Once a Bjorken expansion was employed, we found that our calculated $K/\pi^{+}$ and $(B+\bar{B})/\pi^{+}$ ratios matched experimental data well, which suggests that hadrons do not at all need to start in equilibrium at the onset of the hadron gas phase.


\end{document}